\documentclass[12pt,letterpaper]{article}
\usepackage{graphicx}
\usepackage{amsmath}

\input{tcilatex}

\begin{document}

\author{F. G. Bass$^{1}$\thanks{%
Professor F. G. Bass passed away during preparation of the manuscript.}, and
V. V. Prosentsov$^{2}$\thanks{%
Correspondent author, e-mail: prosentsov@yahoo.com} \\
$^{1}$Ha Pizga 2215, Ariel 40700, Israel\\
$^{2}$Stationsstraat 86, Deurne, 5751 HH, The Netherlands}
\title{Dynamic scattering by cluster of small particles: local perturbation approach}
\maketitle

\begin{abstract}
The wave scattering by moving particles (dynamic scattering) is a well known
physical problem routinely occurring in practice. For the particles which
are much smaller than the incident wavelength, the static scattering problem
can be solved by using the local perturbation method. In this paper we apply
the local perturbation approach to the problem of the dynamic scattering by
the cluster of small particles. We calculate the fields scattered by the
cluster of moving particles. As an example, the scattered field is
calculated for moving sphere in scalar approximation and in vector case.
\end{abstract}

\section{Introduction}

Wave propagation and scattering in inhomogeneous media is a classical
physical problem constantly reoccurring in many practical areas such as
adaptive optics, free space communication, biology, and medicine. In many
practical cases the inhomogeneous medium is actually homogeneous host medium
(infinite or bounded) filled with the finite size inhomogeneities like dust
particles, water droplets, air bubbles, snow flakes, and living cells.

The wave scattering by stationary inhomogeneities (static scattering)\ was
studied extensively, and there are many papers devoted to this problem (see
for example \cite{Kerker}-\cite{Born} and references therein). In reality,
however, some scatterers do move: snow falls, blood cells flow, and cosmic
dust rovers the space.

The wave scattering by moving bodies (dynamic scattering) is a long standing
problem with many practical applications \cite{Bladel}-\cite{DynSc2}. For
example, the scattering properties of the moving particles are routinely
used for velocity and object size measurements \cite{JLee}-\cite{NYokoi}.
The statistical properties of the dynamic scattering are discussed in \cite
{Bladel}, and \cite{Ishimaru}- \cite{Rytov}, while the used scattering
function is essentially of the static particle. The general theory of the
scattering by single tree-dimensional object in translation motion was
presented in \cite{Zutter}, and only recently the exact theory of the
scattering by moving sphere was presented (see the work \cite{Handa} and
references therein). The dynamic scattering by the cluster of particles was
not studied yet.

When the characteristic size of the inhomogeneity is much smaller than the
incident wavelength, the local perturbation method (LPM) can be used. The
LPM was applied initially by Fermi for calculation of atomic spectra \cite
{Fermi}-\cite{Fermi1}. Later, the method was applied in crystal theory \cite
{Kosevich} and solid state physics \cite{Maleev}, \cite{BassFrShefr}. Most
recently, the local perturbation method was applied for wave scattering by
cluster of static particles \cite{BassFix}-\cite{BassFRPros1}.

There are, to authors knowledge, no studies where the LPM was used for study
of the wave propagation in the media filled with moving local perturbations.
The LPM allows, in principle, to take into account multiple scattering by
moving particles, the shape of each moving scatterer, and the resonance
properties of the dynamic scattering.

In this paper we use the LPM to study the dynamic wave scattering by the
cluster of the particles which characteristic sizes are small compared to
the incident wavelength. The general formalism is presented for this
problem. As an example, we apply our method for calculation of the field
scattered by moving sphere in scalar approximation and in vector case.

In the following discussion we will make no distinction between particle and
perturbation.

\section{General formalism: the scattering by the local perturbations moving
with arbitrary speeds}

The wave propagation in the medium filled with the $N$ small particles can
be described by the following equation 
\begin{equation}
\widehat{H}_{0}\left\{ \frac{\partial }{i\partial \mathbf{r}},\frac{%
-\partial }{i\partial t}\right\} \mathbf{E}(\mathbf{r},t)+\sum_{n=1}^{N}%
\widehat{H}_{1}\left\{ \frac{\partial }{i\partial \mathbf{r}},\frac{%
-\partial }{i\partial t}\right\} U_{n}(\mathbf{r}-\mathbf{r}_{n}(t))\times
\label{m1}
\end{equation}

\begin{equation*}
\widehat{H}_{2}\left\{ \frac{\partial }{i\partial \mathbf{r}},\frac{%
-\partial }{i\partial t}\right\} \mathbf{E}(\mathbf{r},t)=\mathbf{j}(\mathbf{%
r},t)
\end{equation*}

where the operators $\widehat{H}_{0}$, $\widehat{H}_{1}$, and $\widehat{H}%
_{2}$ are the tensors of the second order, $\mathbf{E}$ and $\mathbf{j}$ are
the field and source vectors respectively depending on the space and time
coordinates $\mathbf{r}$ and $t$. The function $U_{n}$ describes the
properties of the $n$-th local perturbation and its dimensions, $\mathbf{r}%
_{n}(t)$ is the position of the $n$-th perturbation and this position varies
in time.

We note that the operator $\widehat{H}_{0}$ in the Eq. (\ref{m1}) describes
the field propagation in the homogeneous medium, while the operators $%
\widehat{H}_{1}$ and $\widehat{H}_{2}$ are related to the perturbation.

We emphasize that the Eq. (\ref{m1}) is quite general one and it can be
reduced to partial differential equation, to integral equation, or to
difference equations \cite{BassFRMarad}. As a consequence, the solution of
the Eq. (\ref{m1}) can describe the broad class of the fields related to
different physical phenomena.

In this section we solve the Eq. (\ref{m1}) by using the local perturbation
method. For completeness, we note that the local perturbation method (LPM)
is valid for the particles (perturbations) which characteristic size $L_{n}$
is much smaller compared to the incident wavelength $\lambda $ and that in
this case the following relation holds \cite{BassFix}

\begin{equation}
U_{n}(\mathbf{r}-\mathbf{r}_{n}(t))\mathbf{E}(\mathbf{r},t)\approx U_{n}(%
\mathbf{r}-\mathbf{r}_{n}(t))\mathbf{E}(\mathbf{r}_{n}(t),t),\;(\lambda
/L_{n}\ll 1)  \label{m1a}
\end{equation}
By multiplying the Eq. (\ref{m1}) by the operator $\widehat{H}_{0}^{-1}$
inverse to the operator $\widehat{H}_{0}$\ and by using the LPM relation (%
\ref{m1a}) we can present the field $\mathbf{E}$ in the following form

\begin{equation}
\mathbf{E}(\mathbf{r},t)=\widehat{H}_{0}^{-1}\mathbf{j}(\mathbf{r}%
,t)-\sum_{n=1}^{N}\widehat{H}_{0}^{-1}\widehat{H}_{1}U_{n}(\mathbf{r}-%
\mathbf{r}_{n}(t))\mathbf{F}_{n}(t).  \label{m2}
\end{equation}
Here the field $\mathbf{F}_{n}$ is defined as

\begin{equation}
\mathbf{F}_{n}(t)\equiv \widehat{H}_{2}\mathbf{E}(\mathbf{r}_{n}(t),t)
\label{m2a}
\end{equation}
and $\widehat{H}_{0}^{-1}\widehat{H}_{0}=\widehat{I}$, where $\widehat{I}$
is the unity operator.

The field $\mathbf{E}$ in the Eq. (\ref{m2}) can be presented as the sum of
the incident $\mathbf{E}_{in}$ and the scattered $\mathbf{E}_{sc}$ fields
calculated via the Green's tensors, i. e. as

\begin{equation}
\mathbf{E}(\mathbf{r},t)=\mathbf{E}_{in}(\mathbf{r},t)+\mathbf{E}_{sc}(%
\mathbf{r},t),  \label{m3a}
\end{equation}
where

\begin{eqnarray}
\mathbf{E}_{in}(\mathbf{r},t) &\equiv &\int \widehat{G}_{0}(\mathbf{r}-%
\mathbf{r}^{\prime },t-t^{\prime })\mathbf{j}(\mathbf{r}^{\prime },t^{\prime
})d\mathbf{r}^{\prime }dt^{\prime },\;  \label{m4} \\
\mathbf{E}_{sc}(\mathbf{r},t) &=&\sum_{n=1}^{N}\mathbf{E}_{sc,n},
\label{m4a} \\
\mathbf{E}_{sc,n}(\mathbf{r},t) &\equiv &-\int \widehat{G}_{1}(\mathbf{r}-%
\mathbf{r}^{\prime },t-t^{\prime })U_{n}(\mathbf{r}^{\prime }-\mathbf{r}%
_{n}(t^{\prime }))\mathbf{F}_{n}(t^{\prime })d\mathbf{r}^{\prime }dt^{\prime
}.  \label{m4b}
\end{eqnarray}
Here $\mathbf{E}_{sc,n}$ is the field scattered by the $n$-th particle, $%
\widehat{G}_{0}$ is the Green's tensor of the homogeneous medium, and $%
\widehat{G}_{1}$ is the Green's tensor related to the inhomogeneity

\begin{eqnarray}
\widehat{G}_{0}(\mathbf{r}-\mathbf{r}^{\prime },t-t^{\prime }) &\equiv &%
\frac{\pi ^{-4}}{16}\int \frac{\widehat{H}_{0}^{-1}(\mathbf{q},\omega )}{%
16\pi ^{4}}e^{i\mathbf{q}(\mathbf{r}-\mathbf{r}^{\prime })-i\omega
(t-t^{\prime })}d\mathbf{q}d\omega ,  \label{m5} \\
\widehat{G}_{1}(\mathbf{r}-\mathbf{r}^{\prime },t-t^{\prime }) &\equiv &%
\frac{\pi ^{-4}}{16}\int \widehat{H}_{0}^{-1}(\mathbf{q},\omega )\widehat{H}%
_{1}(\mathbf{q},\omega )e^{i\mathbf{q}(\mathbf{r}-\mathbf{r}^{\prime
})-i\omega (t-t^{\prime })}d\mathbf{q}d\omega .  \label{m5a}
\end{eqnarray}
Here and below we use infinite limits for integration and we do not write
them explicitly. The expressions (\ref{m3a})-(\ref{m5}) allow to calculate
the total field $\mathbf{E}$ in the medium when the fields $\mathbf{F}_{n}$
are known. To find the fields $\mathbf{F}_{n}$ we multiply the Eq. (\ref{m2}%
) by the operator $\widehat{H}_{2}$ and get the following equation for the
fields $\mathbf{F}_{n}$

\begin{equation}
\mathbf{F}_{m}(t)=\mathbf{J}_{m}(\mathbf{r}_{m}(t),t)-\sum_{n=1}^{N}\int 
\widehat{G}_{21}(\mathbf{r}_{m}(t)-\mathbf{r}^{\prime },t-t^{\prime })U_{n}(%
\mathbf{r}^{\prime }-\mathbf{r}_{n}(t^{\prime }))\mathbf{F}_{n}(t^{\prime })d%
\mathbf{r}^{\prime }dt^{\prime },  \label{m7}
\end{equation}
where the vector $\mathbf{J}_{m}$ and the Green's tensor $\widehat{G}_{21}$
are defined as

\begin{eqnarray}
\mathbf{J}_{m}(\mathbf{r}_{m}(t),t) &\equiv &\int \widehat{G}_{2}(\mathbf{r}%
_{m}(t)-\mathbf{r}^{\prime },t-t^{\prime })\mathbf{j}(\mathbf{r}^{\prime
},t^{\prime })d\mathbf{r}^{\prime }dt^{\prime },  \label{m9} \\
\widehat{G}_{21}(\mathbf{r}-\mathbf{r}^{\prime },t-t^{\prime }) &\equiv &%
\frac{\pi ^{-4}}{16}\int \widehat{H}_{2}(\mathbf{q},\omega )\widehat{H}%
_{0}^{-1}(\mathbf{q},\omega )\widehat{H}_{1}(\mathbf{q},\omega )\times 
\notag \\
&&e^{i\mathbf{q}(\mathbf{r}_{m}(t)-\mathbf{r}^{\prime })-i\omega
(t-t^{\prime })}d\mathbf{q}d\omega , \\
\widehat{G}_{2}(\mathbf{r}-\mathbf{r}^{\prime },t-t^{\prime }) &\equiv &%
\frac{\pi ^{-4}}{16}\int \widehat{H}_{2}(\mathbf{q},\omega )\widehat{H}%
_{0}^{-1}(\mathbf{q},\omega )\times  \label{m9a} \\
&&e^{i\mathbf{q}(\mathbf{r}_{m}(t)-\mathbf{r}^{\prime })-i\omega
(t-t^{\prime })}d\mathbf{q}d\omega .  \notag
\end{eqnarray}
We note that the expression (\ref{m7}) is actually the system of equations
with respect to the unknown vectors $\mathbf{F}_{n}$ and it can be presented
in the compact form

\begin{equation}
\sum_{n=1}^{N}\int \widehat{W}_{mn}(t,t^{\prime })\mathbf{F}_{n}(t^{\prime
})dt^{\prime }=\mathbf{J}_{m}(\mathbf{r}_{m}(t),t),  \label{m11}
\end{equation}
where the operators $\widehat{W}_{mn}$ are

\begin{eqnarray}
\widehat{W}_{mm}(t,t^{\prime }) &\equiv &\widehat{I}\delta (t-t^{\prime })+
\label{m13} \\
&&\int \widehat{G}_{21}(\mathbf{r}_{m}(t)-\mathbf{r}_{n}(t^{\prime })-%
\mathbf{r}^{\prime },t-t^{\prime })U_{m}(\mathbf{r}^{\prime })d\mathbf{r}%
^{\prime },  \notag \\
\widehat{W}_{\substack{ mn  \\ (m\neq n)}}(t,t^{\prime }) &\equiv &V_{n}%
\widehat{G}_{21}(\mathbf{r}_{m}(t)-\mathbf{r}_{n}(t^{\prime }),t-t^{\prime
}).  \label{m13a}
\end{eqnarray}
Here $\widehat{I}$ is the unity operator and $V_{n}$ is the volume of the $n$%
-th particle calculated as

\begin{equation}
V_{n}=\int U_{n}(\mathbf{r}^{\prime })d\mathbf{r}^{\prime }.  \label{m13f}
\end{equation}

We note that the fields (\ref{m3a})-(\ref{m4a}) and the fields $\mathbf{F}%
_{n}(t)$ (solutions of the system (\ref{m11})) give complete solution of the
dynamic multiple scattering problem in the local perturbation approximation.

We note also that, the solution of the system (\ref{m11}), in general case,
can not be expressed in analytical form and it should be solved numerically.
However, in particular case when the perturbations move with the constant
speed, the system (\ref{m11}) can be resolved analytically. This solution
will be discussed in the following subsection.

\subsection{The scattering by the local perturbations moving with constant
velocities}

Consider the situation when the perturbations move with constant velocities.
In this case their coordinates $\mathbf{r}_{n}(t)$ are

\begin{equation}
\mathbf{r}_{n}(t)=\mathbf{r}_{0n}+\mathbf{v}_{n}t,\;\left| \mathbf{v}%
_{n}\right| =\text{const, }(\mathbf{v}_{n}\neq \mathbf{v}_{m})  \label{m14}
\end{equation}
where $\mathbf{r}_{0n}$ is the initial position of the $n$-th perturbation
at time $t=0$ and $\mathbf{v}_{n}$ is the velocity of the $n$-th
perturbation. Substituting relation for coordinates (\ref{m14}) into general
expressions (\ref{m13}) for operators $\widehat{W}_{mn}$, we can recast the
system of equations (\ref{m11}) into the following one

\begin{equation}
\widehat{W}_{mm}(\omega )\widetilde{\mathbf{F}}_{m}(\omega )+\sum_{n\neq
m}^{N}\int \widehat{f}_{mn}(\mathbf{q},\omega )\widetilde{\mathbf{F}}%
_{n}(\omega +\mathbf{q(v}_{m}-\mathbf{v}_{n}\mathbf{))}d\mathbf{q}=%
\widetilde{\mathbf{J}}_{m}(\omega ),  \label{m15}
\end{equation}
or in vector components

\begin{equation}
W_{mm,ij}(\omega )\widetilde{F}_{m,j}(\omega )+\sum_{n\neq m}^{N}\int
f_{mn,ij}(\mathbf{q},\omega )\widetilde{F}_{n,j}(\omega +\mathbf{q(v}_{m}-%
\mathbf{v}_{n}\mathbf{))}d\mathbf{q}=\widetilde{J}_{m,i}(\omega ).
\label{m16}
\end{equation}
Here $\widetilde{\mathbf{F}}_{n}$ is the Fourier transform of the field $%
\mathbf{F}_{n}$ and the operator $\widehat{W}_{mn}$ is

\begin{eqnarray}
\widehat{W}_{mm}(\omega ) &=&\widehat{I}+\int \widehat{H}_{2}(\mathbf{q}%
,\omega +\mathbf{qv}_{m})\widehat{H}_{0}^{-1}(\mathbf{q},\omega +\mathbf{qv}%
_{m})\times  \label{m17} \\
&&\widehat{H}_{1}(\mathbf{q},\omega +\mathbf{qv}_{m})\widetilde{U}_{m}(%
\mathbf{q})d\mathbf{q,}  \notag
\end{eqnarray}
and

\begin{eqnarray}
\widehat{f}_{mn}(\mathbf{q},\omega ) &=&\frac{V_{n}}{8\pi ^{3}}\widehat{H}%
_{2}(\mathbf{q},\omega +\mathbf{qv}_{m})\widehat{H}_{0}^{-1}(\mathbf{q}%
,\omega +\mathbf{qv}_{m})\times  \label{m18a} \\
&&\widehat{H}_{1}(\mathbf{q},\omega +\mathbf{qv}_{m})e^{i\mathbf{qr}_{mn}}, 
\notag \\
\mathbf{r}_{mn} &\equiv &\mathbf{r}_{0m}-\mathbf{r}_{0n}.  \label{m18b}
\end{eqnarray}
The Fourier transforms of the source function $\mathbf{J}_{m}(\mathbf{r}%
_{m}(t),t)$ and the function describing the shape of the particle $U_{m}(%
\mathbf{r})$ respectively are

\begin{eqnarray}
\widetilde{\mathbf{J}}_{m}(\omega ) &=&\frac{1}{2\pi }\int \mathbf{J}_{m}(%
\mathbf{r}_{m}(t),t)e^{i\omega t}dt,  \label{m18g} \\
\widetilde{U}_{m}(\mathbf{q}) &=&\frac{1}{8\pi ^{3}}\int U_{m}(\mathbf{r}%
)e^{-i\mathbf{qr}}d\mathbf{r.}  \label{m18gg}
\end{eqnarray}
We note that the expressions (\ref{m15}) and (\ref{m16}) are the system of
equations with respect to the unknown fields $\widetilde{\mathbf{F}}_{n}$,
and even these systems can not be solved analytically without further
simplification.

\subsubsection{Local perturbations moving as one body (all particles have
the same velocity)}

To simplify the systems (\ref{m15}) and (\ref{m16}) further, we assume that
the speeds of the particles are such that the following condition holds

\begin{equation}
\frac{\left| \mathbf{v}_{m}-\mathbf{v}_{n}\right| }{c}\ll 1.  \label{m18h}
\end{equation}
This condition is automatically satisfied for the particles with small
speeds, and it is also correct for the particles with large but similar
speeds. By using the condition (\ref{m18h}), we can approximate the Fourier
transform $\widetilde{\mathbf{F}}_{n}$ as

\begin{equation}
\widetilde{\mathbf{F}}_{n}(\omega +\mathbf{q(v}_{m}-\mathbf{v}_{n}))\approx 
\widetilde{\mathbf{F}}_{n}(\omega )+\left. \frac{\partial \widetilde{\mathbf{%
F}}_{n}(\omega +\mathbf{q(v}_{m}-\mathbf{v}_{n}))}{\partial \omega }\right|
_{\mathbf{v}_{m}=\mathbf{v}_{n}}\mathbf{q(v}_{m}-\mathbf{v}_{n}),
\label{m19a}
\end{equation}
where the second term is much smaller than the first one and it can be
neglected. Neglecting by the second term in Eq. (\ref{m19a}) we effectively
apply condition that all the particles have the same velocity.

Taking into account the relation (\ref{m19a}), we present the system (\ref
{m16}) in the following form

\begin{equation}
\sum_{n=1}^{N}W_{mn,ij}(\omega )\widetilde{F}_{n,j}(\omega )=\widetilde{J}%
_{m,i}(\omega ),  \label{m19b}
\end{equation}
and its solution for the field components $\widetilde{F}_{n,i}$ is

\begin{equation}
\widetilde{F}_{n,i}(\omega )=\sum_{m=1}^{N}\frac{\widehat{A}_{nm,ij}%
\widetilde{J}_{m,j}(\omega )}{\det \widehat{W}(\omega )}.  \label{m21}
\end{equation}
Here the tensor $\widehat{W}$ has components $W_{mn,ij}$ (see the formula (%
\ref{m19b})) and $\widehat{A}_{nm,ij}$ is the matrix of cofactors. Finally,
taking into account the expression (\ref{m21}) for the fields $\widetilde{%
\mathbf{F}}_{n}$, the scattered field (\ref{m4a}) can be presented in the
form

\begin{equation}
\mathbf{E}_{sc}(\mathbf{r},t)=\sum_{n=1}^{N}\mathbf{E}_{sc,n}(\mathbf{r},t),
\label{m40}
\end{equation}
where the filed $\mathbf{E}_{sc,n}$ scattered by the $n$-th particle is

\begin{eqnarray}
\mathbf{E}_{sc,n}(\mathbf{r},t) &=&-\frac{V_{n}}{8\pi ^{3}}\int \widehat{H}%
_{0}^{-1}(\mathbf{q},\omega +\mathbf{qv}_{n})\widehat{H}_{1}(\mathbf{q}%
,\omega +\mathbf{qv}_{n})\times  \label{m40a} \\
&&e^{i\mathbf{q}(\mathbf{r}-\mathbf{r}_{0n}-\mathbf{v}_{n}t)-i\omega t}d%
\mathbf{q}\sum_{m=1}^{N}\frac{\widehat{A}_{nm}\widetilde{\mathbf{J}}%
_{m}(\omega )}{\det \widehat{W}(\omega )}d\omega .  \notag
\end{eqnarray}

Furthermore, we note that the field (\ref{m40}) can be integrated over $%
\omega $ space by using the residue theorem and in this case the scattered
field is

\begin{equation}
\mathbf{E}_{sc}(\mathbf{r},t)=\int \frac{\mathbf{Q}(\omega ,t)e^{-i\omega t}%
}{\det \widehat{W}(\omega )}d\omega =2\pi i\sum_{q}\frac{\mathbf{Q}(\omega
_{q},t)e^{-i\omega _{q}t}}{\left. \frac{d\det \widehat{W}(\omega )}{d\omega }%
\right| _{\omega =\omega _{q}}},  \label{m50}
\end{equation}
where the vector $\mathbf{Q}$ is defined as

\begin{eqnarray}
\mathbf{Q}(\omega ,t) &\equiv &-\sum_{n,m=1}^{N}\frac{V_{n}}{8\pi ^{3}}%
\widehat{A}_{nm}\widetilde{\mathbf{J}}_{m}(\omega )\int \widehat{H}_{0}^{-1}(%
\mathbf{q},\omega +\mathbf{qv}_{n})\times  \label{m55} \\
&&\widehat{H}_{1}(\mathbf{q},\omega +\mathbf{qv}_{n})e^{i\mathbf{q}(\mathbf{r%
}-\mathbf{r}_{0n}-\mathbf{v}_{n}t)}d\mathbf{q.}  \notag
\end{eqnarray}
Here $\omega _{q}$ is the $q$-th root of the equation $\det \widehat{W}%
(\omega )=0$. Furthermore, we note that the resonance frequencies of the
dynamic scattering are defined by the equation

\begin{equation}
\det \widehat{W}(\omega )=0.  \label{m57}
\end{equation}

We note that the formula (\ref{m50}) is the essence of this subsection. The
formula gives analytical expression for the field scattered by the particles
moving with the same speed in the local perturbation approximation.

\section{Example 1: Scattering by moving sphere in scalar approximation}

In this section we consider the scattering by moving sphere in scalar
approximation. We assume that the particle moves in the infinite homogeneous
medium with the constant velocity $\mathbf{v}$ in $x$ direction, and that
the radius and the volume of the sphere is $L$ and $V$ respectively. The
position of the sphere is described by the radius vector $\mathbf{r}_{1}(t)=%
\mathbf{r}_{01}+\mathbf{v}t$, and $\mathbf{r}_{01}$ is the position of the
particle at time $t=0$. In this case, the equation for the scalar field $E(%
\mathbf{r},t)$ is

\begin{equation}
\left( \Delta -\frac{\varepsilon _{h}}{c^{2}}\frac{\partial ^{2}}{\partial
t^{2}}\right) E(\mathbf{r},t)-\frac{(\varepsilon _{sc}-\varepsilon _{h})}{%
c^{2}}\frac{\partial ^{2}}{\partial t^{2}}U(\mathbf{r}-\mathbf{r}%
_{1}(t))E_{1}(\mathbf{r}_{1},t)=j(\mathbf{r},t),  \label{m60}
\end{equation}
where $\varepsilon _{h}$ and $\varepsilon _{sc}$ are the permittivities of
the host medium and the particle respectively, $U$ is the function
describing the shape of the sphere. Comparing Eq. (\ref{m60}) with the
general Eq. (\ref{m1}) we can see that the operators $\widehat{H}_{0}$, $%
\widehat{H}_{1}$, and $\widehat{H}_{2}$ are

\begin{eqnarray}
\widehat{H}_{0}\left\{ \frac{\partial }{i\partial \mathbf{r}},\frac{%
-\partial }{i\partial t}\right\} &=&\Delta -\frac{\varepsilon _{h}}{c^{2}}%
\frac{\partial ^{2}}{\partial t^{2}},\;\widehat{H}_{2}\left\{ \frac{\partial 
}{i\partial \mathbf{r}},\frac{-\partial }{i\partial t}\right\} =1,
\label{m65} \\
\widehat{H}_{1}\left\{ \frac{\partial }{i\partial \mathbf{r}},\frac{%
-\partial }{i\partial t}\right\} &=&-\frac{(\varepsilon _{sc}-\varepsilon
_{h})}{c^{2}}\frac{\partial ^{2}}{\partial t^{2}}  \label{m65a}
\end{eqnarray}
and as the result

\begin{eqnarray}
\widehat{H}_{0}\left\{ \mathbf{q},\omega \right\} &=&-q^{2}+k^{2},\;k\equiv 
\sqrt{\varepsilon _{h}}\frac{\omega }{c},  \label{m67} \\
\widehat{H}_{1}\left\{ \mathbf{q},\omega \right\} &=&\frac{(\varepsilon
_{sc}-\varepsilon _{h})\omega ^{2}}{c^{2}},\;\widehat{H}_{2}\left\{ \mathbf{q%
},\omega \right\} =1.  \label{m67a}
\end{eqnarray}
By using the obtained results (\ref{m40a}) and the expressions (\ref{m67})-(%
\ref{m67a}) we get for the scattered field $E_{sc}$ the following expression

\begin{equation}
E_{sc}(\mathbf{r},t)=\frac{(\varepsilon _{sc}-\varepsilon _{h})V}{8\pi
^{3}c^{2}}\int \frac{(\omega +\mathbf{qv})^{2}e^{i\mathbf{q}(\mathbf{r}-%
\mathbf{r}_{01})-i(\omega +\mathbf{qv})t}}{q^{2}-\left( k+\sqrt{\varepsilon
_{h}}\frac{\mathbf{qv}}{c}\right) ^{2}}d\mathbf{q}\widetilde{E}_{1}(\omega
)d\omega ,  \label{m70}
\end{equation}
where $\widetilde{E}_{1}(\omega )$ is the field inside the particle and it is

\begin{equation}
\widetilde{E}_{1}(\omega )=\widetilde{E}_{inc,1}(\omega )+\frac{(\varepsilon
_{sc}-\varepsilon _{h})}{c^{2}}\widetilde{E}_{1}(\omega )\int \frac{%
\widetilde{U}(\mathbf{q})(\omega +\mathbf{qv})^{2}}{q^{2}-\left( k+\sqrt{%
\varepsilon _{h}}\frac{\mathbf{qv}}{c}\right) ^{2}}d\mathbf{q.}  \label{m72}
\end{equation}
We note that the integral in Eq. (\ref{m70}) can be calculated with the help
of the stationary phase method for the large distances when $kR\gg 1$ ($%
R\equiv \left| \mathbf{r}-\mathbf{r}_{1}(t)\right| $). Integrating both
formulae (\ref{m70}) and (\ref{m72}) over $\mathbf{q}$ we get for the
scattered field and the field inside particle $\widetilde{E}_{1}(\omega )$
the following expressions respectively

\begin{align}
E_{sc}(\mathbf{r},t)& =\frac{(\varepsilon _{sc}-\varepsilon _{h})V}{4\pi
c^{2}}\gamma (\mathbf{r},t)\int \omega ^{2}\widetilde{E}_{1}(\omega
)e^{i\omega (\varphi (t)-t)}d\omega ,\;(kR\gg 1)  \label{m72aa} \\
\widetilde{E}_{1}(\omega )& =\frac{\widetilde{E}_{inc,1}(\omega )}{W(\omega )%
},\;R\equiv \left| \mathbf{r}-\mathbf{r}_{1}(t)\right| ,  \label{m72ab}
\end{align}
where the coefficients $\gamma $ and $\varphi $ are

\begin{eqnarray}
\gamma (\mathbf{r},t) &\equiv &\frac{\left( 1+\beta \frac{R_{x}(t)}{\rho }%
\right) ^{2}}{(1-\beta ^{2})\rho (t)},\;\rho (t)\equiv \sqrt{R^{2}-\beta
^{2}R_{\perp }^{2}},\;\beta \equiv \sqrt{\varepsilon _{h}}\frac{v}{c},
\label{m73} \\
\varphi (t) &\equiv &\frac{\sqrt{\varepsilon _{h}}\rho (t)}{c\left( 1-\beta
^{2}\right) }\left( 1+\beta \frac{R_{x}(t)}{\rho (t)}\right)
,\;R_{x}(t)\equiv \sqrt{R^{2}-R_{\perp }^{2}},  \label{m74} \\
R_{\bot } &\equiv &\left| \mathbf{r}_{\bot }-\mathbf{r}_{1\bot }(t)\right|
,\;v=\left| \mathbf{v}\right| ,
\end{eqnarray}
and the denominator $W$ is

\begin{equation}
W(\omega )=1-(\varepsilon _{sc}-\varepsilon _{h})\left\{ 
\begin{array}{c}
\frac{1}{\varepsilon _{h}}\left( \ln (\frac{1+\beta }{1-\beta })/2\beta
-1\right) \\ 
+\frac{\omega ^{2}L^{2}}{2c^{2}}\frac{1}{\left( 1-\beta ^{2}\right) ^{2}} \\ 
+i\frac{\omega ^{3}L^{3}}{3c^{3}}\frac{\sqrt{\varepsilon _{h}}\left( 1+\beta
^{2}\right) }{\left( 1-\beta ^{2}\right) ^{3}}
\end{array}
\right\} .  \label{m75}
\end{equation}
We note that the formula (\ref{m75}) is correct even for the relatively high
velocities when $\beta >0.1$. For the static particles ($\beta =0$), the
formula (\ref{m75}) reproduces well known result presented, for example, in 
\cite{BassFRPros1}.

\subsection{The resonance}

The formula (\ref{m75}) shows that the dynamic scattering in the scalar case
has resonance when

\begin{equation}
\func{Re}W(\omega _{r})=0.  \label{m79f}
\end{equation}
From the resonance condition (\ref{m79f}) we can calculate the resonance
frequency of the field scattered by the moving sphere in scalar approximation

\begin{equation}
\omega _{r}=\frac{\sqrt{2}c(1-\beta ^{2})}{L\sqrt{\varepsilon
_{sc}-\varepsilon _{h}}}\left[ 1-\frac{(\varepsilon _{sc}-\varepsilon _{h})}{%
\varepsilon _{h}}\left( \ln \left( \frac{1+\beta }{1-\beta }\right) /2\beta
-1\right) \right] ^{1/2}.  \label{m80}
\end{equation}
The expression (\ref{m80}) clearly shows that the resonance frequency $%
\omega _{r}$ decreases with the speed of the particle and that the resonance
frequency can be even zero. Moreover, the higher the optical contrast of the
particle, the faster decrease of the frequency (see the \ref{Fig1}).

The expression (\ref{m80}) can be simplified for small speeds when $\beta
\ll 1$, and in this case the resonance frequency $\omega _{r}$ of the field
scattered by the moving particle is

\begin{equation}
\omega _{r}=\frac{\sqrt{2}c}{L\sqrt{\varepsilon _{sc}-\varepsilon _{h}}}%
\left( 1-(\varepsilon _{sc}-\varepsilon _{h})\frac{\beta ^{2}}{3\varepsilon
_{h}}\right) ^{1/2},\;\left( \beta \ll 1\right)  \label{m81}
\end{equation}
and the resonance width $\xi $ is

\begin{equation}
\xi \equiv \left. \frac{\func{Im}W}{\frac{\partial \func{Re}W}{\partial
\omega }}\right| _{\omega =\omega _{r}}=\frac{2c\sqrt{\varepsilon _{h}}}{%
9L(\varepsilon _{sc}-\varepsilon _{h})}\frac{\left( 3+\beta ^{2}\right) }{%
(1-\beta ^{2})^{3}}\left( 1-(\varepsilon _{sc}-\varepsilon _{h})\frac{\beta
^{2}}{3\varepsilon _{h}}\right) .  \label{m85}
\end{equation}
We note that the resonance frequency and the resonance width are the
functions of the particle's speed $v$. At zero speed when $\beta =0$, the
formula (\ref{m81}) reproduces the result obtained previously for the
resonance scattering by static particle \cite{BassFRPros1}. Here assumed
that the refractive indexes of the particle and the host medium are real
values.

The formula (\ref{m81}) shows that the resonance frequency decreases with
the speed of the particle (we consider the most commonly encountered case
when $\varepsilon _{sc}>\varepsilon _{h}$), and for particles with
relatively high speeds the resonance frequency may be even zero. Physically
this means that light propagating inside particle with the speed about $c/%
\sqrt{\varepsilon _{sc}}$ does not interact with boundaries of the particle
moving with the speed $v$.

The resonance width (\ref{m85}) is more complicated function of the
particle's speed: it can increase or decrease its value at some conditions.
For the small speeds when $\beta \ll 1$, we have

\begin{equation}
\xi \approx \frac{2c\sqrt{\varepsilon _{h}}}{3L(\varepsilon
_{sc}-\varepsilon _{h})}\left( 1+\frac{\beta ^{2}}{3}\left( 11-\frac{%
\varepsilon _{sc}}{\varepsilon _{h}}\right) \right) ,\;(\beta \ll 1)
\label{m85b}
\end{equation}
meaning that the width increases with the increase of the particle's speed
when $\varepsilon _{sc}<11\varepsilon _{h}$. On the contrary, when $%
\varepsilon _{sc}>11\varepsilon _{h}$, the resonance width decreases with
the increase of the speed of the particle when $\beta \ll 1$.

\subsection{The scattered intensities}

The scattered field (\ref{m72aa}) can be calculated even further when the
incident field $\widetilde{E}_{inc,1}$ is somehow specified. Consider two
most common cases below.

\subsubsection{Case 1: monochromatic incident light}

Suppose that the incident field is a monochromatic light with the angular
frequency $\Omega $. In this case the incident field can be presented in the
following form

\begin{equation}
\widetilde{E}_{inc,1}(\omega )=E_{1}\delta (\omega -\Omega ),  \label{m86}
\end{equation}
where $E_{1}$ is the amplitude of the field and $\delta $ is the delta
function. In accordance with (\ref{m72aa}) and (\ref{m86}) the expression
the scattered field and its intensity $I_{sc}\equiv \left| E_{sc}(\mathbf{r}%
,t)\right| ^{2}$ is

\begin{eqnarray}
E_{sc}(\mathbf{r},t) &=&\frac{V}{4\pi }(\varepsilon _{sc}-\varepsilon _{h})%
\frac{\Omega ^{2}}{c^{2}}\frac{\gamma (\mathbf{r},t)E_{1}}{W(\Omega )}%
e^{i\Omega (\varphi (t)-t)},  \label{m87} \\
I_{sc}(\mathbf{r},t) &=&\frac{V^{2}}{16\pi ^{2}}\left| \varepsilon
_{sc}-\varepsilon _{h}\right| ^{2}\frac{\Omega ^{4}}{c^{4}}\frac{\gamma ^{2}(%
\mathbf{r},t)\left| E_{1}\right| ^{2}}{\left| W(\Omega )\right| ^{2}}.
\label{m87a}
\end{eqnarray}
The formula (\ref{m87a}) shows that the intensity of the scattered field
vary in space and time via the coefficient $\gamma (\mathbf{r},t)$. The
intensity increases when the particle heads in the direction of observer and
it goes down when the particle flies away from the observer. The scattered
intensity is maximal then the frequency of the incident light $\Omega $
coincides with the resonance frequency $\omega _{r}$ of the field scattered
by the moving particle (Eq. (\ref{m81})), because in this case the
denominator $W$ is minimal.

\subsubsection{Case 2: broad band light}

Suppose now that the incident field is relatively broad function in
frequency domain and that the resonance frequency $\omega _{r}$ of the
particle is inside this frequency band. In this case the integral in (\ref
{m72aa}) can be calculated with the help of the residue theorem and we get
the following expressions for the scattered field and its intensity

\begin{eqnarray}
E_{sc}(\mathbf{r},t) &=&i\frac{V}{2}(\varepsilon _{sc}-\varepsilon _{h})%
\frac{\omega _{0}^{2}}{c^{2}}\gamma (\mathbf{r},t)\frac{\widetilde{E}%
_{inc,1}(\omega _{0})}{\left. \frac{\partial W}{\partial \omega }\right|
_{\omega =\omega _{0}}}e^{i\omega _{0}(\varphi (t)-t)},  \label{m88} \\
E_{sc}(\mathbf{r},t) &=&\frac{V^{2}}{4}\left| \varepsilon _{sc}-\varepsilon
_{h}\right| ^{2}\frac{\left| \omega _{0}^{2}\right| ^{2}}{c^{4}}\gamma ^{2}(%
\mathbf{r},t)\left| \frac{\widetilde{E}_{inc,1}(\omega _{0})}{\left. \frac{%
\partial W}{\partial \omega }\right| _{\omega =\omega _{0}}}\right| ^{2}e^{-2%
\func{Im}[\omega _{0}(\varphi (t)-t)]},  \label{m88a}
\end{eqnarray}
where $\omega _{0}$ is the solution of the equation $W(\omega )=0$ (see Eq. (%
\ref{m75})) and $\omega _{0}$ is, in principle, complex number. The formula (%
\ref{m88a}) for the intensity of the scattered field is correct when the
condition

\begin{equation}
\func{Im}[\omega _{0}(\varphi (t)-t)]\geq 0  \label{m90}
\end{equation}
is satisfied. The formula (\ref{m88a}) shows that the intensity of the
scattered field decreases exponentially for the times which are not equal to 
$t=\varphi (t)$.

When the resonance is narrow, $\func{Im}\omega _{0}=-\xi $ meaning that the
intensity (\ref{m88a}) decrease is related to the resonance width: the
broader the resonance the faster the scattered intensity drops. In the
limit, when the resonance width tends to zero, the scattered intensity does
not decay exponentially in time.

\section{Example 2: Scattering by moving sphere in vector case}

In this section we consider the resonance scattering by moving sphere in
vector case. As well as in the scalar case, we assume that the particle
moves in the infinite homogeneous medium with the constant velocity $\mathbf{%
v}$ in $x$ direction, and that the radius and the volume of the sphere is $L$
and $V$ respectively. The position of the sphere is described by the radius
vector $\mathbf{r}_{1}(t)=\mathbf{r}_{01}+\mathbf{v}t$, where $\mathbf{r}%
_{01}$ is the position of the particle at time $t=0$. In this case, the
equation for the vector field $\mathbf{E}(\mathbf{r},t)$ is

\begin{equation}
\left( \Delta -\mathbf{\nabla }\otimes \mathbf{\nabla -}\frac{\varepsilon
_{h}}{c^{2}}\frac{\partial ^{2}}{\partial t^{2}}\right) \mathbf{E}(\mathbf{r}%
,t)-\frac{(\varepsilon _{sc}-\varepsilon _{h})}{c^{2}}\frac{\partial ^{2}}{%
\partial t^{2}}U(\mathbf{r}-\mathbf{r}_{1}(t))\mathbf{E}_{1}(\mathbf{r}%
_{1},t)=\mathbf{j}(\mathbf{r},t).  \label{m160}
\end{equation}
Here $\Delta $ and $\mathbf{\nabla }$\ are the Laplacian and nabla
operators, $\otimes $ defines tensor product, $\varepsilon _{h}$ and $%
\varepsilon _{sc}$ are the permittivities of the host medium and the
particle respectively, $U$ is the function describing the shape of the
sphere. Comparing Eq. (\ref{m160}) with the general Eq. (\ref{m1}) we can
see that the operators $\widehat{H}_{0}$, $\widehat{H}_{1}$, and $\widehat{H}%
_{2}$ are

\begin{eqnarray}
\widehat{H}_{0}\left\{ \frac{\partial }{i\partial \mathbf{r}},\frac{%
-\partial }{i\partial t}\right\} &=&\Delta -\mathbf{\nabla }\otimes \mathbf{%
\nabla }-\frac{\varepsilon _{h}}{c^{2}}\frac{\partial ^{2}}{\partial t^{2}}%
,\;\widehat{H}_{2}\left\{ \frac{\partial }{i\partial \mathbf{r}},\frac{%
-\partial }{i\partial t}\right\} =1,  \label{m165} \\
\widehat{H}_{1}\left\{ \frac{\partial }{i\partial \mathbf{r}},\frac{%
-\partial }{i\partial t}\right\} &=&-\frac{(\varepsilon _{sc}-\varepsilon
_{h})}{c^{2}}\frac{\partial ^{2}}{\partial t^{2}}  \label{m165a}
\end{eqnarray}
and as the result the operators $\widehat{H}_{0}$, $\widehat{H}_{1}$, and $%
\widehat{H}_{2}$ are

\begin{eqnarray}
\widehat{H}_{0}\left\{ \mathbf{q},\omega \right\} &=&-q^{2}+\mathbf{q}%
\otimes \mathbf{q}+k^{2},\;k\equiv \sqrt{\varepsilon _{h}}\frac{\omega }{c},
\label{m167} \\
\widehat{H}_{1}\left\{ \mathbf{q},\omega \right\} &=&\frac{(\varepsilon
_{sc}-\varepsilon _{h})\omega ^{2}}{c^{2}},\;\widehat{H}_{2}\left\{ \mathbf{q%
},\omega \right\} =1.  \label{m167a}
\end{eqnarray}
By using the obtained results (\ref{m40a}) and the expressions (\ref{m167})-(%
\ref{m167a}) we get for the scattered field $\mathbf{E}_{sc}$ the following
expression

\begin{eqnarray}
\mathbf{E}_{sc}(\mathbf{r},t) &=&\frac{(\varepsilon _{sc}-\varepsilon _{h})V%
}{8\pi ^{3}c^{2}}\int \left( \widehat{I}-\frac{\mathbf{q}\otimes \mathbf{q}}{%
\left( k+\sqrt{\varepsilon _{h}}\frac{\mathbf{qv}}{c}\right) ^{2}}\right) 
\widetilde{\mathbf{E}}_{1}(\omega )d\omega \times  \label{m170} \\
&&\frac{(\omega +\mathbf{qv})^{2}e^{i\mathbf{q}(\mathbf{r}-\mathbf{r}%
_{01})-i(\omega +\mathbf{qv})t}}{q^{2}-\left( k+\sqrt{\varepsilon _{h}}\frac{%
\mathbf{qv}}{c}\right) ^{2}}d\mathbf{q},  \notag
\end{eqnarray}
where $\widetilde{\mathbf{E}}_{1}(\omega )$ is the field inside the particle
and it is

\begin{eqnarray}
\widetilde{\mathbf{E}}_{1}(\omega ) &=&\widetilde{\mathbf{E}}_{inc,1}(\omega
)+\frac{(\varepsilon _{sc}-\varepsilon _{h})}{c^{2}}\times  \label{m172} \\
&&\int \left( \widehat{I}-\frac{\mathbf{q}\otimes \mathbf{q}}{\left( k+\sqrt{%
\varepsilon _{h}}\frac{\mathbf{qv}}{c}\right) ^{2}}\right) \frac{\widetilde{%
\mathbf{E}}_{1}(\omega )\widetilde{U}(\mathbf{q})(\omega +\mathbf{qv})^{2}}{%
q^{2}-\left( k+\sqrt{\varepsilon _{h}}\frac{\mathbf{qv}}{c}\right) ^{2}}d%
\mathbf{q.}  \notag
\end{eqnarray}
We note that the integral in Eq. (\ref{m170}) can be calculated with the
help of the stationary phase method for the large distances when $kR\gg 1$ ($%
R\equiv \left| \mathbf{r}-\mathbf{r}_{1}(t)\right| $).

Integrating both formulae (\ref{m170}) and (\ref{m172}) over $\mathbf{q}$ we
get for the scattered field and the field inside particle $\widetilde{%
\mathbf{E}}_{1}(\omega )$ the following expressions respectively

\begin{align}
\mathbf{E}_{sc}(\mathbf{r},t)& =\frac{(\varepsilon _{sc}-\varepsilon _{h})}{%
4\pi c^{2}}V\left[ \gamma (\mathbf{r},t)\int \omega ^{2}\widetilde{\mathbf{E}%
}_{1}(\omega )e^{i\omega (\varphi (t)-t)}d\omega \right. +  \label{m172aa} \\
& \left. \frac{(1-\beta ^{2})c^{2}}{\varepsilon _{h}}\mathbf{\nabla }\otimes 
\mathbf{\nabla }\frac{1}{\rho (t)}\int \widetilde{\mathbf{E}}_{1}(\omega
)e^{i\omega (\varphi (t)-t)}d\omega \right] ,  \notag \\
R& \equiv \left| \mathbf{r}-\mathbf{r}_{1}(t)\right| ,\;(kR\gg 1) \\
\widehat{D}(\omega )\widetilde{\mathbf{E}}_{1}(\omega )& =\widetilde{\mathbf{%
E}}_{inc,1}(\omega ),\;D_{ij}=\delta _{ij}W_{j},  \label{m172ab}
\end{align}
where the coefficients $\gamma $ and $\varphi $ are explained in formulae (%
\ref{m73})-(\ref{m74}), and the coefficients $W_{j}$ are

\begin{eqnarray}
W_{x}(\omega ) &=&1-(\varepsilon _{sc}-\varepsilon _{h})\left\{ 
\begin{array}{c}
\frac{1}{\varepsilon _{0}}\left[ \frac{1-\beta ^{2}}{\beta ^{2}}+\frac{\beta
^{2}-1}{2\beta ^{3}}\ln \left( \frac{1+\beta }{1-\beta }\right) \right] + \\ 
\frac{L^{2}\omega ^{2}}{2c^{2}}\left[ \frac{1}{\beta ^{2}(1-\beta ^{2})}-\ln
\left( \frac{1+\beta }{1-\beta }\right) /2\beta ^{3}\right] + \\ 
i\frac{2\omega ^{3}L^{3}}{9c^{3}}\frac{\sqrt{\varepsilon _{h}}}{\left(
1-\beta ^{2}\right) ^{2}}
\end{array}
\right\} ,  \label{m175} \\
W_{y,z}(\omega ) &=&1-(\varepsilon _{sc}-\varepsilon _{h})\left\{ 
\begin{array}{c}
\frac{1}{\varepsilon _{0}}\left[ -\frac{1+2\beta ^{2}}{2\beta ^{2}}+\frac{%
\beta ^{2}+1}{4\beta ^{3}}\ln \left( \frac{1+\beta }{1-\beta }\right) \right]
+ \\ 
\frac{L^{2}\omega ^{2}}{2c^{2}}\left[ \frac{3\beta ^{2}-1}{2\beta
^{2}(1-\beta ^{2})^{2}}+\ln \left( \frac{1+\beta }{1-\beta }\right) /4\beta
^{3}\right] + \\ 
i\frac{2\omega ^{3}L^{3}}{9c^{3}}\frac{\sqrt{\varepsilon _{h}}\left( 1+\beta
^{2}\right) }{\left( 1-\beta ^{2}\right) ^{3}}
\end{array}
\right\} ,  \label{m176} \\
\beta &\equiv &\sqrt{\varepsilon _{h}}\frac{v}{c}.
\end{eqnarray}
We note that the formulae (\ref{m175}) and (\ref{m176}) are correct even for
the relatively high velocities when $\beta >0.1$. For the static particle,
the expressions (\ref{m175}) and (\ref{m176}) transform to the known values
presented, for example, in \cite{BassFRPros1}.

\subsection{The resonance}

The formulae (\ref{m175}) and (\ref{m176}) suggest that the dynamic
scattering in vector case has two resonances defined by the two following
equations

\begin{equation}
\func{Re}W_{x}(\omega _{r,x})=0,\;\func{Re}W_{y,z}(\omega _{r,yz})=0.
\label{m180}
\end{equation}
The resonance frequencies are

\begin{eqnarray}
\omega _{r,x} &=&\frac{\sqrt{2}c}{L\sqrt{\varepsilon _{sc}-\varepsilon _{h}}}%
\sqrt{\frac{1-\frac{(\varepsilon _{sc}-\varepsilon _{h})}{\varepsilon
_{h}\beta ^{2}}\left[ 1-\beta ^{2}+(\beta ^{2}-1)\ln \left( \frac{1+\beta }{%
1-\beta }\right) /2\beta \right] }{\frac{1}{(1-\beta ^{2})\beta ^{2}}-\ln
\left( \frac{1+\beta }{1-\beta }\right) /2\beta ^{3}}}  \label{m180f} \\
\omega _{r,yz} &=&\frac{\sqrt{2}c}{L\sqrt{\varepsilon _{sc}-\varepsilon _{h}}%
}\sqrt{\frac{1+\frac{(\varepsilon _{sc}-\varepsilon _{h})}{2\varepsilon
_{h}\beta ^{2}}\left[ 1+2\beta ^{2}-(\beta ^{2}+1)\ln \left( \frac{1+\beta }{%
1-\beta }\right) /2\beta \right] }{\frac{3\beta ^{2}-1}{2(1-\beta
^{2})^{2}\beta ^{2}}+\ln \left( \frac{1+\beta }{1-\beta }\right) /4\beta ^{3}%
}}  \label{m180h}
\end{eqnarray}
The obtained expressions for the resonance frequencies (\ref{m180f}) and (%
\ref{m180h}) are not transparent due to complex relations between $\beta $
and logarithmic function. For the small speeds when $\beta \ll 1$, the
resonance frequencies $\omega _{r}$ of the field scattered by the moving
particle are

\begin{eqnarray}
\omega _{r,x} &=&\omega _{r,0}\left( 1-\varsigma \beta ^{2}/5\right)
,\;\omega _{r,yz}=\omega _{r,0}\left( 1-2\varsigma \beta ^{2}/5\right) ,
\label{m181} \\
\omega _{r,0} &\equiv &\frac{c}{L}\sqrt{\frac{2\varepsilon _{h}+\varepsilon
_{sc}}{\varepsilon _{h}(\varepsilon _{sc}-\varepsilon _{h})}},\varsigma
\equiv \frac{(4\varepsilon _{sc}+5\varepsilon _{h})}{(\varepsilon
_{sc}+2\varepsilon _{h})},\;(\beta \ll 1).
\end{eqnarray}
We do not consider the resonance width here, because the resonance is broad
even for the static particle (see, for example \cite{BassFRPros1}).

The expressions (\ref{m180f})-(\ref{m181}) show that as well as in the
scalar case, the resonance frequencies decrease with the speed of the
particle (we assumed that $\varepsilon _{sc}>\varepsilon _{h}$). However, in
distinction to the scalar case, there are two resonance frequencies of the
scattered field in the vector case: in the direction of the particle
propagation and in the perpendicular direction. In addition, the formulae (%
\ref{m181}) shows that ratio of the frequencies $\omega _{r,x}/\omega
_{r,yz} $ grows with the particle's speed as

\begin{equation}
\frac{\omega _{r,x}}{\omega _{r,yz}}=1+\varsigma \beta ^{2}/5.  \label{m185}
\end{equation}

We note, that the scattered intensities can be calculated in the same way as
it was done for the scalar case, and we will not do it here.

\section{Conclusions}

The new method describing the wave propagation and scattering in the medium
filled with the small moving particle has been proposed. The explicit
analytical solution was presented for the field scattered by the particles
moving with the constant speed.

As an example, the field scattered by the small moving sphere is studied. It
was shown that in the scalar case, the speed of the particle changes the
resonance width and essentially affects the decay rate of the scattered \
intensity. It was shown also that in the vector case, the resonance
frequency is different in direction of movement and in the direction
transverse to the movement.

\begin{equation*}
\end{equation*}

\textbf{Acknowledgment}

We would like to thank Prof. V. Freilikher for critical comments and
important suggestions.

\begin{equation*}
\end{equation*}

\end{document}